\documentclass[aps,prl,twocolumn,superscriptaddress]{revtex4-2}
\usepackage{graphicx}
\usepackage{dcolumn}
\usepackage{bm}
\usepackage{amsmath}
\usepackage{xcolor}
\usepackage{lineno}

\begin{document}


\title{Tunable superconducting diode effect in higher-harmonic InSb nanosheet interferometers}
\author
{Xingjun Wu,\textsuperscript{1, \textsection} Ji-Yin Wang,$^{1}$ Haitian Su,$^{2,3}$ Shili Yan,$^{1}$\\
Dong Pan,$^{4, \ast}$ Jianhua Zhao,$^{4}$ Po Zhang,$^{1, \dag}$ and H. Q. Xu$^{1, 2, \ddag}$\\
\normalsize{$^{1}$Beijing Academy of Quantum Information Sciences, Beijing 100193, China}\\
\normalsize{$^{2}$Beijing Key Laboratory of Quantum Devices, Key Laboratory for the Physics and Chemistry}\\
\normalsize{of Nanodevices, and School of Electronics, Peking University, Beijing 100871, China}\\
\normalsize{$^{3}$Institute of Condensed Matter and Material Physics, School of Physics, Peking University,}\\
\normalsize{Beijing 100871, China}\\
\normalsize{$^{4}$State Key Laboratory of Superlattices and Microstructures, Institute of Semiconductors,}\\
\normalsize{Chinese Academy of Sciences, P.O. Box 912, Beijing 100083, China}\\
\normalsize{\textsuperscript{\textsection}  wuxj@baqis.ac.cn, $^\ast$  pandong@semi.ac.cn, $^\dag$  zhangpo@baqis.ac.cn, $^\ddag$ hqxu@pku.edu.cn}
}

\begin{abstract}
Superconducting diodes, characterized by the nonreciprocal supercurrent flow, have gained significant attention for their potential in dissipationless electronics. This study presents a superconducting quantum interference device (SQUID) composed of two Al-InSb nanosheet Josephson junctions. Utilizing prepatterned local backgates, we achieve a gate- and flux-tunable superconducting diode with controllable efficiency in both amplitude and sign. Numerical simulations attribute the diode effect to higher harmonics in the current-phase relation. Crucially, fractional Shapiro step experiments provide direct insights into the evolution of these higher harmonics with flux tuning, showcasing significant enhancements in the second-harmonic signatures of the SQUID near half-integer flux quanta. Furthermore, we investigate the microwave-assisted diode response and experimentally show that the polarity of the diode effect can be switched by the microwave power. These results demonstrate the potential of InSb nanosheet-based hybrid devices
as highly tunable elements for use in dissipationless electronics.
\end{abstract}
\maketitle

\par The superconducting diode effect (SDE) is a nonreciprocal transport phenomenon, analogous to charge transport in semiconductor $p-n$ junctions, where dissipationless superconducting currents flow in one direction while being blocked or partially blocked in the opposite direction. This unique characteristic positions superconducting diodes as promising elements for applications in superconducting electronics, spintronics, and quantum technologies\cite{braginski2019superconductor,wendin2017quantum,linder2015superconducting,cai2023superconductor}. The SDE has recently attracted worldwide attention and has been the subject of study in various superconducting systems, including Josephson junctions (JJs) and junction-free superconductors\cite{ando2020observation,mazur2022gate,itahashi2020nonreciprocal,lin2022zero,baumgartner2022supercurrent,turini2022josephson,wu2022field,de2021gate,diez2023symmetry,trahms2023diode,sundaresh2023diamagnetic,matsuo2023josephson,lotfizadeh2024superconducting,PhysRevLett.133.087001}. The microscopic mechanisms underlying the SDE involve intrinsic material properties, including finite-momentum pairing\cite{pal2022josephson,yuan2022supercurrent,davydova2022universal}, the magneto-chiral effect\cite{legg2022superconducting}, and valley polarization\cite{PhysRevLett.130.266003}, as well as extrinsic factors like trapped vortices and self-inductance\cite{suri2022non,sundaresh2023diamagnetic, PhysRevB.6.855}.

\par For JJs, the current-phase relation (CPR) is typically expressed as a Fourier series, $I = \sum\limits_{n}I_{n}sin(n\varphi)$, where $\varphi$ is the phase drop across the junction and $I_{n}$ represents the current component of the $n$th harmonic term. A purely sinusoidal CPR results in equal amplitudes of positive and negative critical currents, thus precluding the SDE. However, high-transparency JJs generally possess a non-sinusoidal CPR, and in such higher-harmonic JJs ($n \textgreater 1$), a non-zero phase difference $\delta$ between harmonics can induce SDE, as illustrated by expressions like $I = I_{1}sin(\varphi)+I_{2}sin(2\varphi+\delta)$\cite{zhang2024evidencephi0josephsonjunctionskewed,PhysRevLett.130.266003}. Achieving such non-zero $\delta$ typically requires breaking time-reversal and inversion symmetries, which can arise from an interplay of external magnetic field with spin-orbit coupling or a nonuniform potential\cite{PhysRevB.93.155406,PhysRevB.89.195407,turini2022josephson,mazur2022gate}.  

\par Nonreciprocal critical currents in SQUIDs have been recognized since the 1970s\cite{fulton1970current,PhysRevB.6.855}, but they have recently gained renewed attention due to the involvement of high-order harmonics in CPR\cite{ciaccia2023gate,ciaccia2024charge,valentini2024parity,gupta2023gate,larsen2020parity,PhysRevResearch.6.033281,10.21468/SciPostPhys.16.1.030,PhysRevB.106.134514,souto2022josephson,reinhardt2024link}. Theoretical proposals suggest that higher-harmonic CPR in SQUIDs with highly transmissive JJs can significantly affect supercurrent nonreciprocity, with predictions of a butterfly pattern diagram to illustrate diode efficiency\cite{souto2022josephson,PhysRevB.106.134514}. These theoretical frameworks set the stage for experimental investigations into SDE in various hybrid systems with SQUID configurations. Recent experiments have focused on quantum well-based hybrid superconductor-semiconductor interferometers, such as those made from Al-InAs and Al-Ge quantum wells, where gate-tunable $sin(2\varphi)$ elements are anticipated to enable parity-protected qubits\cite{ciaccia2024charge,valentini2024parity,PhysRevResearch.6.033281,gupta2023gate,ciaccia2023gate,larsen2020parity}. Tuning these diode responses with a microwave field is predicted to yield intriguing results, such as achieving an ideal SDE\cite{souto2022josephson,valentini2024parity} or even switching the polarity of the diode\cite{PhysRevResearch.6.023011,PhysRevResearch.6.L022002}, the latter of which has yet to be reported by experiments. These results offer valuable insights for creating a highly tunable superconducting diode.

InSb nanosheet is a two-dimensional (2D) semiconductor nanostructure developed in recent years\cite{de2016twin,pan2016nl}. With its strong spin-orbit interaction, small effective mass, large Land$\acute{e}$ $g$-factor\cite{de2016twin,nilsson2009giant,lei2021gate}, and 2D geometry, InSb nanosheets have emerged as a highly promising platform for the development of novel planar hybrid quantum devices\cite{PhysRevResearch.1.032031,Kang2019nl,turini2022josephson,iorio2023half}. While prior studies have predominantly focused on devices incorporating single JJs, including the intrinsic diode effect in InSb nanosheets \cite{turini2022josephson}, research on nanosheet-based interferometers remains limited. The development of such interferometers holds the potential to introduce additional tuning knobs, facilitating the exploration of novel hybrid quantum devices on nanosheets.


In our work, we engineered the SDE in an InSb nanosheet interferometer with higher-harmonic junctions. As opposed to the intrinsic diode in single JJs\cite{turini2022josephson}, the SDE in a SQUID does not rely on spin-orbit coupling or in-plane Zeeman fields. Instead, tuning the two asymmetric SQUID arms enables a wider range of control over diode rectification efficiency. In the experiment, we demonstrate effective control of the supercurrent in each junction using two pre-patterned local backgates. This enables the realization of a gate- and flux-tunable superconducting diode with a controllable diode efficiency, both in amplitude (up to $\sim$10\%) and direction. We analyze the mechanism and find that higher-order Josephson harmonics in CPR are indispensable for the SDE. We utilize fractional Shapiro steps to monitor the higher harmonic content in flux-tuning experiments. Notably, we observe significant enhancements in the second-harmonic signatures of the SQUID near half-integer flux quanta. Furthermore, we investigate the microwave response and experimentally demonstrate that the SDE can not only be tuned by microwave power, but also be switched. This may open up new possibilities for application with a microwave-controlled diode effect\cite{PhysRevResearch.6.023011}. Numerical calculations qualitatively support all observations.

\begin{figure}[t] 
\centering
\includegraphics[width=1\linewidth] {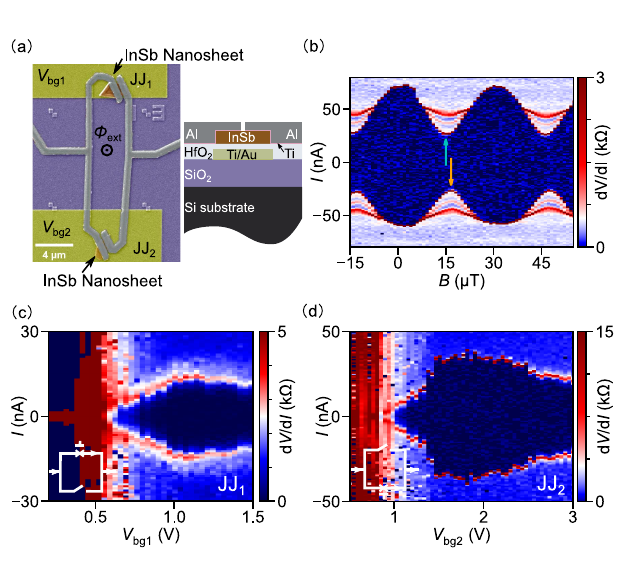}
\caption{Device and characterization. (a) False-color scanning electron microscopy image of Al-InSb SQUID and the cross-section of the Al-InSb nanosheet JJ that makes up the SQUID. The brown regular polygon regions are two free-standing InSb nanosheets. (b) Differential resistance d$V$/d$I$ of the SQUID as a function of bias current $I$ and perpendicular magnetic field $B$ at $V_{bg1}$ = 1.069~V and $V_{bg2}$ = 1.8~V. The cyan and orange arrows illustrate the extreme positions of the switching and retrapping valleys respectively. (c) and (d) Differential resistance d$V$/d$I$ as a function of bias current and backgate voltage for JJ$_{1}$ and JJ$_{2}$. When one junction is measured, the other is completely pinched off by the corresponding backgate voltage.} \label{fig1}
\end{figure}

\par Figure~\ref{fig1}(a) shows a false-color scanning electron microscope image of an Al-InSb interferometer. Two planar JJs (JJ$_{1}$ and JJ$_{2}$), each consisting of Al and InSb nanosheets with a gap of about 100~nm, are looped by the Al superconducting wires to form an SQUID (see Supplemental Material for a zoomed-in image of JJ$_{1}$ and JJ$_{2}$\cite{supp}). The free-standing InSb nanosheets are grown by molecular-beam epitaxy, with a typical thickness of 10 to 100~nm. Details about the material and transport properties have been studied in previous reports\cite{pan2016nl,Kang2019nl,wu2024superconductingswitchingjumpinduced}. A set of prepatterned metallic (Ti/Au) backgates are fabricated beneath the InSb nanosheets, isolated by a 15~nm HfO$_{2}$ dielectric layer. They are used to independently tune the superconductivity of JJ$_{1}$ and JJ$_{2}$ by applying respective backgate voltages. InSb nanosheets are mechanically transferred by a tip, followed by a standardized aluminum evaporation process. Fabrication details are presented in Supplementary Material\cite{supp}. A standard DC measurement is performed in a dilution refrigerator with a base temperature of about 15~mK. 

\par In Fig.~\ref{fig1}(c) and \ref{fig1}(d), we present the differential resistance of JJ$_{i}$ ($i$ = 1, 2) as a function of the respective backgate voltage $V_{bgi}$ and bias current $I$. When measuring one JJ, a backgate voltage of 0~V is applied to pinch off the other JJ. The critical current of JJ$_{1}$ can be tuned from a pinch-off status at $V_{bg1}\leq$ 0.5~V to approximately 15~nA at $V_{bg1}$ = 1.07~V, while the critical current of JJ$_{2}$ can be tuned from pinch-off at $V_{bg2}\leq$ 0.9~V up to 40~nA at $V_{bg2}$ = 1.8~V. A slight decrease in the critical current $I_{c}$ is observed as $V_{bg}$ continues to increase. This could be related to the distribution of carriers at high $V_{bg}$. At high $V_{bg}$ values, carriers are pushed away from the superconductor-semiconductor interface due to electrostatic potential effects, thereby affecting the superconducting proximity effect and reducing $I_{c}$. Additionally, excessive carrier accumulation at high $V_{bg}$ may enhance electron-electron interactions and scattering, which could also contribute to the suppression of $I_{c}$. Similar non-monotonic behavior of $I_{c}$ with $V_{bg}$ has been reported in other JJ systems\cite{yan2023supercurrent,li2020fermi}. Compared to JJ$_{1}$, the stochastic switching current in JJ$_{2}$ exhibits a more pronounced response to thermal and charge fluctuations when tuning $V_{bg2}$, as illustrated in Fig.~\ref{fig1}(d).

\par
\begin{figure}[t] 
\centering
\includegraphics[width=1\linewidth] {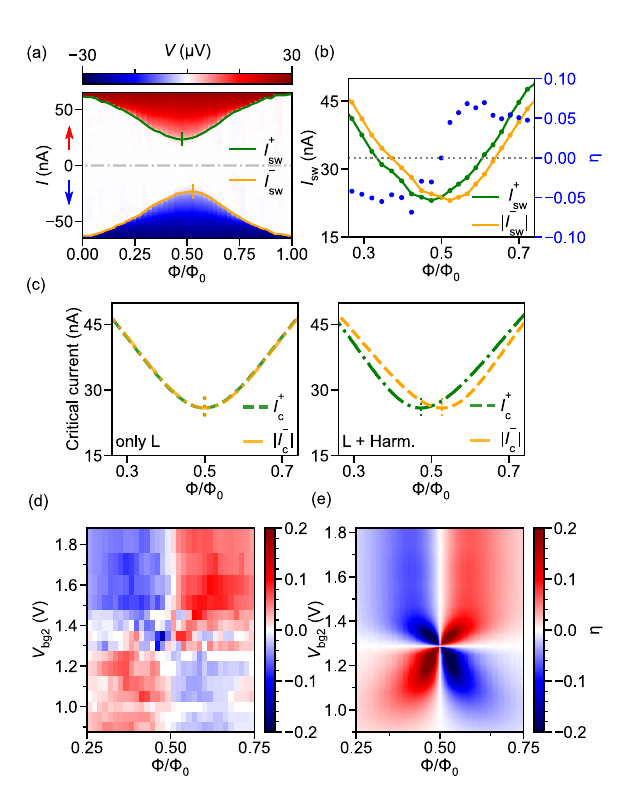}
\caption{Gate- and flux-tunable SDE. (a) The measured voltage $V$ as a function of $I$ and $\Phi/\Phi_{0}$ at a gate configuration of ($V_{bg1},V_{bg2}$) = (1.069~V, 1.819~V). $I$ is swept from zero to positive values (indicated by the red arrow) in the red branch, and subsequently from zero to negative values (indicated by the blue arrow) in the blue branch, allowing the extraction of switching currents $I_{sw}^{+}$ and $I_{sw}^{-}$ for two opposite sweep directions. (b) $I_{sw}^{+}$ and $|I_{sw}^{-}|$, extracted from panel (a), plotted as a function of $\Phi/\Phi_{0}$. The blue dotted line corresponds the diode efficiency. (c) $I_{c}^{+}$ and $|I_{c}^{-}|$, simulated using two different parameters, as a function of $\Phi/\Phi_{0}$. Parameters in the left panel consider only a realistic $L$ = 100~pH, while parameters in the right include $L$ = 100~pH and the first two harmonics ($K_{1}$, $K_{2}$) = (0.96, 0.04). Parameters ($I_{sw1}$, $I_{sw2}$) = (18~nA, 44~nA), obtained at a gate configuration of ($V_{bg1}$,$V_{bg2}$) = (1.069~V, 1.819~V), are used for both plots. (d) and (e) Efficiency $\eta$ as a function of $V_{bg2}$ and $\Phi/\Phi_{0}$, respectively, obtained from the experiment and theoretical model. Here, $V_{bg1}$ is fixed at 1.069~V. In panel (d), the pixel distribution along the $V_{bg2}$ axis is non-uniform due to unequal spacing in $V_{bg2}$ sampling.}
\label{fig2}
\end{figure}

 Figure~\ref{fig1}(b) shows the differential resistance of the SQUID as a function of perpendicular magnetic field $B$ and bias current $I$ at a gate configuration of $V_{bg1}$ = 1.069~V and $V_{bg2}$ = 1.8~V, where $I$ is swept from negative to positive in the `+' direction. Three main observations are as follows: (1) The SQUID oscillation is periodic with a period of 32~$\mu$T. This periodicity corresponds to an added external flux quantum $\Phi_{0}$, which gives the area of the inner loop $S$ = 64.7~$\mu$m$^{2}$. This value is slightly larger than the actual inner area of our SQUID ($S$ = 55~$\mu$m$^{2}$). This discrepancy could be due to the magnetic flux focusing effect or London penetration. (2) A set of fringes emerges in the oscillation valley. The discrete fringes observed may arise from Andreev bound states (ABSs) within the induced superconducting gap or from multiple Andreev reflections. As the critical current at the valley decreases, the finite voltage step at the superconducting transition drops below the bias voltage necessary to access the characteristic energy levels within the subgap, thereby facilitating the emergence of these fringes. (3) An asymmetry between the switching and retrapping branches is observed, representing an indirect signature of SDE. Overheating, however, can lead to an artificial SDE: once the junction switches to the resistive state, it fails to return to the superconducting state due to thermal effects until $I$ decreases below the smaller retrapping current, $I_{re}$, resulting in $I_{re}$ \textless $I_{sw}$. To exclude this artificial SDE caused by overheating, a standard approach in diode experiments involves comparing the switching current in the `+' direction ($I_{sw}^{+}$) with that in the opposite `--' direction ($I_{sw}^{-}$), as illustrated in Fig.~\ref{fig2}(a). 

\par To quantify SDE, we study $I_{sw}^{+}$ and $I_{sw}^{-}$ obtained in two opposite sweep directions. The red branch in Fig.~\ref{fig2}(a) corresponds to the current sweeping from zero to a positive value in the `+' direction, while the blue branch corresponds to sweeping from zero to a negative value in the `--' direction. An asymmetric configuration of the SQUID is established in the measurement using $V_{bg1}$ and $V_{bg2}$, with $I_{sw1}$ = 18~nA and $I_{sw2}$ = 44~nA (see Supplemental Material for the determination method of $I_{sw1}$ and $I_{sw2}$ in the SQUID\cite{supp}). $I_{sw}^{+}$ and $|I_{sw}^{-}|$ as functions of the external magnetic flux $\Phi$ are extracted and shown as green and orange curves, respectively, in Fig.~\ref{fig2}(b). A clear asymmetry in the bias current direction is observed, indicating the presence of SDE. The polarity of the SDE changes around $\Phi = 0.5\Phi_{0}$, as reflected in the diode efficiency $\eta$, defined as
\begin{equation} \eta = \frac{\left|I_{sw}^{+}\right| - \left|I_{sw}^{-}\right|}{\left|I_{sw}^{+}\right| + \left|I_{sw}^{-}\right|} \label{eq:6} \end{equation}

\par Next, we attempt to understand the origin of the SDE in this experiment. Recognizing that loop inductance can result in an SDE in SQUIDs, we model an inductive SQUID following Ref.~\citenum{tesche1977dc}. The total current $I_{tot}$ passing through the SQUID is the sum of the currents through JJ$_1$ and JJ$_2$:
\begin{equation}
    I_{tot}(\varphi_{1},\varphi_{2}) = I_{1}(\varphi_{1}) + I_{2}(\varphi_{2}) \label{eq:1}
\end{equation}
Here $\varphi_{1}$ and $\varphi_{2}$ represent the phase drops across JJ$_{1}$ and JJ$_{2}$, respectively. The loop inductance $L$ introduces additional flux into the system, with its magnitude directly proportional to the circulating current $J$ in the loop:
\begin{gather}
    J=(I_{1}(\varphi_{1})-I_{2}(\varphi_{2}))/2 \label{eq:2} \\
    \varphi_{2}-\varphi_{1}=2\pi\Phi_{ext}/\Phi_{0}+2\pi LJ/\Phi_{0} \label{eq:3}
\end{gather} 
where $\Phi_{ext}$ indicates a flux induced by the external magnetic field and $\Phi_{0}=h/2e$ represents the flux quantum. We disregard inductance asymmetry between branches as they are negligible in our SQUID setup. Given a specific $\Phi_{ext}$, $I_{tot}^{+}$ or $I_{tot}^{-}$ can be determined by identifying the maximum or minimum $I_{tot}$ concerning $\varphi_{1}$. 

\par To analyze the impact of inductance on the SDE, we first assume a standard sinusoidal CPR for JJ$_{1}$ and JJ$_{2}$. i.e., $I_{1}=I_{sw1}sin\varphi_{1}$, $I_{2}=I_{sw2}sin\varphi_{2}$. A realistic $L$ in SQUID is the sum of the geometric and kinetic inductances, here estimated to be 100~pH (see Supplemental Material for details\cite{supp}). The left panel in Fig.~\ref{fig2}(c) shows a numerical fit with $L$ = 100~pH. We find that, simply including a realistic $L$ does not adequately capture the asymmetry in the switching current observed in the experiment. To address this asymmetry, $L$ should be at least an order of magnitude larger than the realistic value. Thus, the contribution of higher-order harmonics to the SDE is significant in our SQUID, even if the higher harmonic content is relatively small. For simplicity, we assume that JJ$_{1}$ and JJ$_{2}$ exhibit the same harmonic content, leading to the CPR:
\begin{gather}
    I_{1}(\varphi_{1})=\alpha_{1} I_{sw1}\sum_{n}(-1)^{n+1}K_{n}sin(n\varphi_{1}) \label{eq:4} \\
    I_{2}(\varphi_{2})=\alpha_{2} I_{sw2}\sum_{n}(-1)^{n+1}K_{n}sin(n\varphi_{2}) \label{eq:5}
\end{gather}
Here, $K_{n}$ represents the $n$th harmonic content, and $\alpha_{1,2}$ is a normalized coefficient ensuring $max[I_{1}(\varphi_{1})]=I_{sw1}$ and $max[I_{2}(\varphi_{2})]=I_{sw2}$. To simplify the analysis, we focus on the first two harmonics in this study. The right panel in Fig.~\ref{fig2}(c) illustrates a numerical fit with $K_{1}$ = 0.96, $K_{2}$ = 0.04 and $L$ = 100~pH. Incorporating contributions from the first two harmonics, with a small $K_{2}$, effectively captures crucial experimental characteristics.  

\par We now return to quantify the SDE and discuss its dependence on gate voltage and external flux. Our measurements demonstrate that the flux-dependent SDE can be tuned with varying gate voltages (see Supplementary Material for details\cite{supp}). To systematically quantify the SDE, we extract $\eta$ as a function of $V_{bg2}$ and external flux $\Phi/\Phi_{0}$ at $V_{bg1}$ = 1.069~V, as depicted in Fig.~\ref{fig2}(d). The key observation is that both the magnitude and the sign of $\eta$ can be tuned by $V_{bg2}$ and $\Phi/\Phi_{0}$. When $V_{bg2}$ $\textgreater$ 1.3~V, $\eta$ $\textless$ 0 for $\Phi/\Phi_{0}$ $\textless$ 0.5 and $\eta$ $\textgreater$ 0 for $\Phi/\Phi_{0}$ $\textgreater$ 0.5. Conversely, when $V_{bg2}$ $\textless$ 1.3~V, the scenario is reversed. The experiment yields a maximum $\left|\eta\right|$ of around 10$\%$. Numerical simulation results in Fig.~\ref{fig2}(e) show $\eta$ as a function of $V_{bg2}$ and $\Phi/\Phi_{0}$. In the simulation, we assume the CPR in JJ$_{i}$ ($i$ = 1, 2) remains in the form of $I_{i}=\alpha I_{swi}[K_{1}sin(\varphi_{i})-K_{2}sin(2\varphi_{i})]$, with $K_{1}$ = 0.96, $K_{2}$ = 0.04, and $I_{swi}$ as a gate-dependent switching current in JJ$_{i}$. The experiment and numerical simulation highlight a critical value of $V_{bg2}$ $\approx$ 1.3~V, where $\eta$ changes sign, corresponding to a symmetric SQUID case ($I_{sw1}$ = $I_{sw2}$). The absence of asymmetry between $I_{sw1}$ and $I_{sw2}$ leads to a decrease in $\left|\eta\right|$. However, in practical experiments, especially when $I_{sw1}$ and $I_{sw2}$ are nearly equal, fluctuations in diode polarity are more likely to be triggered by the stochastic nature of $I_{sw}$, as seen at $V_{bg2}$ $\approx$ 1.3~V in Fig.~\ref{fig2}(d).

\par The agreement between the experiment results and the theoretical model highlights the critical role of the high-order harmonic term in CPR. While the CPR of single junctions is not measured due to the absence of a reference junction with a significantly larger critical current, the nonsinusoidal CPR in single JJs inevitably contributes to the emergence of high-order harmonics in the SQUID. These harmonics can be effectively investigated through the AC Josephson effect. In an SQUID without high-order harmonics, quantized voltage plateaus (Shapiro steps) appear at $V$ = $N\frac{hf}{2e}$, with $N$ being an integer, $h$ the Planck constant and $f$ the microwave frequency. When the SQUID includes a high-order harmonic term, such as $sin(n\varphi)$ with $n$ $\textgreater$ 1, fractional quantized voltage plateaus emerge at $V$ = $\frac{N}{n}\frac{hf}{2e}$. Thus, by monitoring the evolution of these fractional Shapiro steps in flux-tuning experiments, one can gain direct insights into the behavior of the relevant high-order harmonic term in the SQUID.
\begin{figure*}[t]
\includegraphics[width=1\linewidth]
{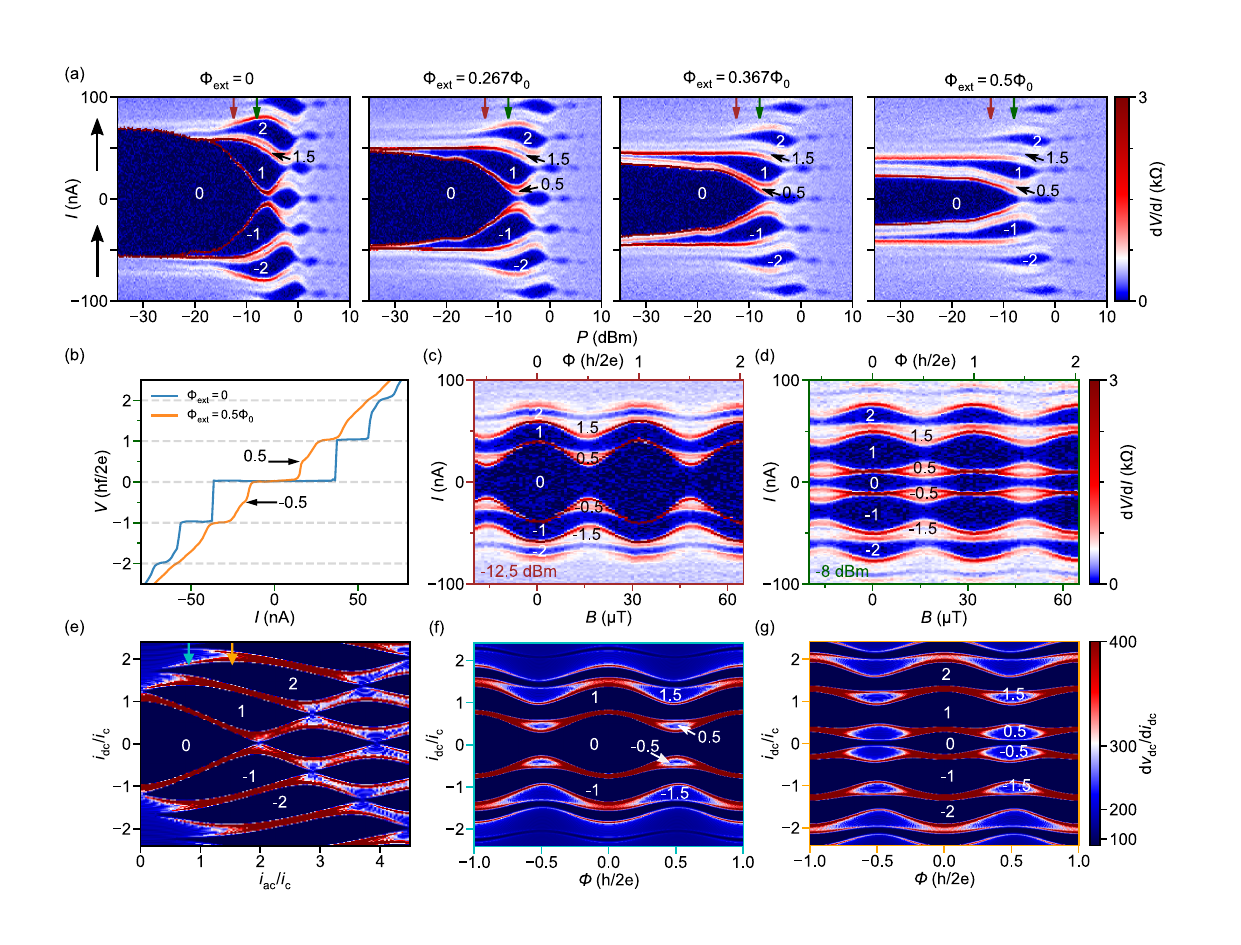}
\caption{Enhanced second-harmonic signatures near half-integer flux quanta. (a) Differential resistance d$V$/d$I$ as a function of $I$ and $P$ when $\Phi_{ext}$ is tuned from 0 to 0.5$\Phi_{0}$. The microwave frequency is 6~GHz, and the gate configuration is $V_{bg1}$ = 1.069~V and $V_{bg2}$ = 1.8~V. The black arrow on the axis indicates the sweeping direction of the dc current bias. The top brown and darkgreen arrows point to the power values of $P$ =-12.5 and -8~dBm. (b) Voltage-current curves at $\Phi_{ext}$ = 0 and 0.5$\Phi_{0}$, as extracted from panel (a) at $P$ = -12.5~dBm. Voltage is normalized to $hf/2e$. Half-integer Shapiro steps appear at $\Phi_{ext}$ = 0.5$\Phi_{0}$, highlighted by the black arrows. (c-d) Differential resistance d$V$/d$I$ versus $I$ and $\Phi$ at $P$ = -12.5~dBm and -8~dBm. The integer and half-integer Shapiro step regions have been labeled, with half-integer steps primarily observed near half-integer flux quanta. (e) Calculated differential resistance map as a function of the dc and ac current bias at $\Phi_{ext}$ = 0. Here, a RSJ model is used, disregarding Joule overheating and nonlinear resistance effects. At $i_{ac}/i_{c}$ = 0.806 (cyan arrow) and 1.53 (orange arrow), the differential resistance d$v_{dc}$/d$i_{dc}$ versus $i_{dc}$ and $\Phi$ is plotted in panel (f) and (g), respectively. Parameters used for
all simulation plots are $R$ =300~$\Omega$, $K_{1}$ = 0.96, $K_{2}$ = 0.04, $f$ = 6~GHz.}
\label{fig3}
\end{figure*}

\begin{figure*}[t]
\includegraphics[width=1\linewidth]
{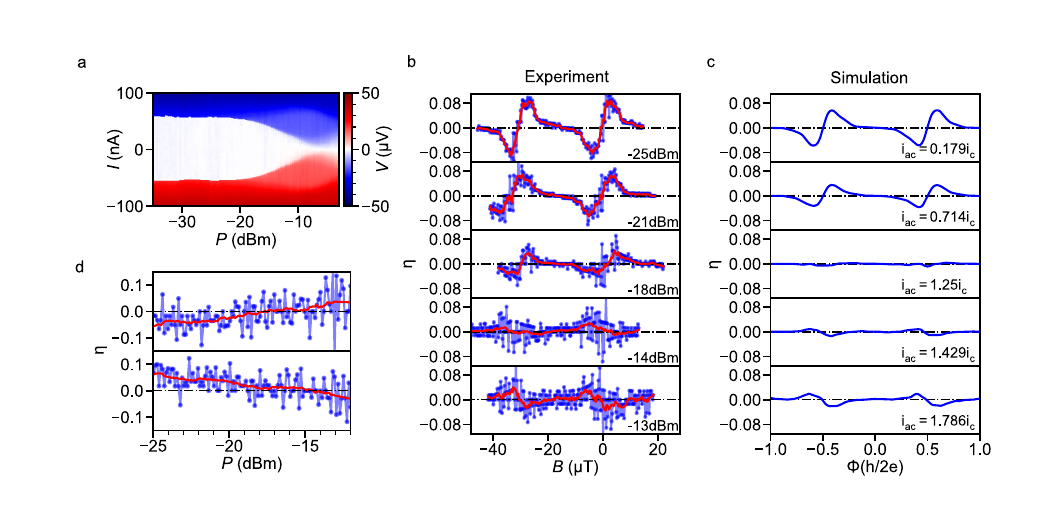}
\caption{Tuning diode nonreciprocity with the microwave power at 8~GHz. (a) The voltage drop V across the SQUID as a function of $P$ and $I$ at zero $B$. (b) The experimental diode efficiency $\eta$ as a function of $B$ at different $P$. $B$ is recalibrated for each line by subtracting offsets. (c) The numerically simulated efficiency $\eta$ as a function of $\Phi$ with varying $i_{ac}$. (d) Efficiency $\eta$ as a function of $P$ at a fixed $\Phi$, with $\Phi$ $\textless$ 0.5$\Phi_{0}$ in the top panel and $\Phi$ $\textgreater$ 0.5$\Phi_{0}$ in the bottom panel. The red curves in panel (b) and (d), fitted to $\eta$, are obtained using a Savitzky-Golay filter.}
\label{fig4}
\end{figure*}

\par In the following, we study the microwave response of our system. The microwave is coupled to the sample via an antenna positioned approximately 1 mm away. Due to frequency-dependent coupling and the lack of impedance matching between the antenna and the sample, accurately quantifying the microwave power reaching the sample is challenging. Therefore, as is common in similar experiments, we report only the nominal microwave power output from the generator at room temperature. 

\par Figure~\ref{fig3}(a) shows differential resistance d$V$/d$I$ as a function of microwave power $P$ and bias current $I$ at different $\Phi_{ext}$. The microwave driving frequency is 6~GHz. The minima in the d$V$/d$I$ curves correspond to the Shapiro steps in the $V$-$I$ characteristic. The low-order integer Shapiro step regions are marked in Fig.~\ref{fig3}(a). Three main observations are as follows: (1) In addition to the integer Shapiro steps, half-integer steps are also observed at zero $\Phi_{ext}$, appearing as dips between adjacent integer Shapiro step lobes. This indicates the presence of higher-order harmonics in our SQUID. (2) As $\Phi_{ext}$ is tuned from 0 to 0.5$\Phi_{0}$, the first half-integer Shapiro step region appears between the first Shapiro step lobes of index 0 and $\pm$1. At the same time, the zero-voltage region shrinks significantly. These behaviors can also be directly identified by comparing the $V$-$I$ curves at $\Phi_{ext}/\Phi_{0}$ = 0 and 0.5 under the same microwave power. As shown in Fig.~\ref{fig3}(b), when fixing $P$ = -12.5~dBm, a weakly developed quantized voltage step at $V = \frac{1}{2}\frac{hf}{2e}$ emerges at $\Phi_{ext}=0.5\Phi_{0}$, while a sharp voltage jump occurs instead at $\Phi_{ext}=0$. (3) Higher-order half-integer Shapiro step regions are expanded by tuning $\Phi_{ext}$ to 0.5$\Phi_{0}$. For instance, in the area around  (-10~dBm, 55~nA) in Fig.~\ref{fig3}(a), which corresponds to the Shapiro step of index 3/2, there is a notable vertical expansion of the Shapiro step region as $\Phi_{ext}$ is tuned from 0 to 0.5$\Phi_{0}$. 

\par Studying the $\Phi_{ext}$-dependent Shapiro steps at fixed $P$ provides a comprehensive understanding of the effect of external flux on higher-order harmonics. Figure~\ref{fig3}(c) and \ref{fig3}(d) show the cases of $P$ = -12.5 and -8~dBm, respectively. We observe, first, that in Fig.~\ref{fig3}(c) and \ref{fig3}(d), the Shapiro step islands, characterized by an index of 0.5, become apparent at half-integer values of $\Phi_{ext}/\Phi_{0}$. These islands gradually disappear once $\Phi_{ext}/\Phi_{0}$ deviates from half-integer values. Second, we observe a periodic oscillation of the Shapiro step regions with indices 3/2 and 5/2 as a function of $\Phi_{ext}$, with a period of $\Phi_{0}$. These half-integer Shapiro step regions reach maxima at half-integer values of $\Phi_{0}$ and minima at integer values of $\Phi_{0}$. 

\par To analyze the dynamics of the SQUID in the microwave field, we consider a resistively shunted junction (RSJ) model with a current bias, disregarding Joule overheating and nonlinear resistance effects. The time-dependent phase difference $\varphi(t)$ and the voltage drop $V(t) = (\hbar/2e)\dot{\varphi}(t)$ are given by
\begin{equation}
    \frac{\hbar}{2eR}\dot{\varphi}(t) + I[\varphi(t)] = i_{dc} + i_{ac}sin(\omega t) \label{eq:7}
\end{equation}
Where $i_{dc}$ and $i_{ac}$
are the amplitudes of the dc and ac amplitudes, $\omega$ is the microwave frequency, $R$ is a shunt resistance in parallel with the interferometer and $I[\varphi(t)]$ is the total CPR of the interferometer, which, as mentioned above, considers only the first two harmonics. By solving this equation numerically, we find that, despite the simplicity of the model, the results qualitatively capture the main features of this work. Figure~\ref{fig3}(e) presents the simulated Shapiro pattern at $\Phi_{ext}/\Phi_{0}$ = 0 and at a drive frequency of 6~GHz. At $i_{ac}/i_{c}$ = 0.806 and 1.53, indicated by the cyan and orange arrows, the calculated differential resistances d$v_{dc}$/d$i_{dc}$ versus $\Phi$ and $i_{dc}$ are depicted in Fig.~\ref{fig3}(f) and 
~\ref{fig3}(g). These results agree well with the main features observed in Fig.~\ref{fig3}(c) and \ref{fig3}(d). 

Next, we discuss the reasons behind the evolution of the half-integer Shapiro steps. Given that the CPR in JJ$_{i}$ ($i$ = 1, 2) follows $I_{JJ_{i}}=\alpha I_{ci}[K_{1}sin(\varphi)-K_{2}sin(2\varphi)]$, the CPR of the SQUID at $\Phi_{ext}$ = 0 and 0.5$\Phi_{0}$ thus can be given by $$I_{\Phi_{ext}=0}=\alpha (I_{c1}+I_{c2})[K_{1}sin(\varphi)-K_{2}sin(2\varphi)]$$ $$I_{\Phi_{ext}=0.5\Phi_{0}}=\alpha (I_{c2}-I_{c1})K_{1}sin(\varphi)-\alpha (I_{c2}+I_{c1})K_{2}sin(2\varphi)$$

By tuning $\Phi_{ext}$, the ratio between the second and first harmonics increases from $\left|\frac{K_{2}}{K_{1}}\right|$ at $\Phi_{ext}$ = 0 to $\left|\frac{I_{c2}+I_{c1}}{I_{c2}-I_{c1}}\right|\left|\frac{K_{2}}{K_{1}}\right|$ at $\Phi_{ext}=0.5\Phi_{0}$. The enhanced second-harmonic content influences the behavior of the half-integer Shapiro steps near $\Phi_{ext}=0.5\Phi_{0}$. In the symmetric limit where $I_{c1}=I_{c2}=I_{c}$, the first harmonic can be completely suppressed at $\Phi_{ext}=0.5\Phi_{0}$, with the second harmonic ($I_{\Phi_{ext}}=-2\alpha K_{2}I_{c}sin(2\varphi)$) contributes to the supercurrent. This regime corresponds to the transport of pairs of Cooper pairs. However, the relatively small magnitude of the second harmonics makes the direct observation of the second harmonic-induced supercurrent component particularly challenging. In addition, we have noticed that in Fig. 3, under the influence of DC bias, magnetic flux, and microwave power, we did not observe the pronounced asymmetric Shapiro step formation or the displaced Shapiro steps phenomenon that is typically seen in JJ array structures\cite{10.1063/5.0121779} or asymmetric SQUID ratchets\cite{PhysRevLett.77.2292,souto2022josephson}. The absence of such clear asymmetry is related to the relatively low content of higher-order harmonic terms in the CPR of the SQUID (see Supplemental Material for a simulation with a higher $\left|\frac{K_{2}}{K_{1}}\right|$).

\par In the final section of this study, we delve into the diode response by tuning microwave powers. The theoretical investigations have revealed that as power levels increase, the efficiency $\eta$ can potentially reach 100$\%$\cite{souto2022josephson,PhysRevResearch.6.L022002,clarke2006squid}. This ideal diode has been experimentally reported\cite{valentini2024parity,PhysRevLett.133.087001}. A further theoretical study recently revealed that tuning $i_{ac}$ may also switch the diode polarity\cite{PhysRevResearch.6.L022002,PhysRevResearch.6.023011}, but this has not yet been experimentally reported. 

\par In Fig.~\ref{fig4}(a), we measure the voltage drop $V$ across the SQUID under microwave irradiation with a drive frequency of $f$ = 8~GHz. As the microwave power $P$ increases, the switching current is gradually suppressed. At a constant $P$, we measure the $V-I$ curves under different $B$ by sweeping the current in two opposite directions, thus obtaining $I_{sw}^{+}$ and $I_{sw}^{-}$ as functions of $B$. Consequently, we determine $\eta$ by utilizing $I_{sw}^{+}$ and $I_{sw}^{-}$, and plot them against $B$ in Fig.~\ref{fig4}(b). Despite the presence of noise in the signal, it is clear that the diode polarity, in addition to being periodically influenced by $B$ (or magnetic flux $\phi$), is also governed by $P$. A switch in diode polarity occurs when $P$ is around -14~dBm. The reproducibility of this result can be partially validated by observing similar behavior over two consecutive magnetic flux quantum periods. At the same time, we notice that the fluctuations in $\eta$ become more pronounced at elevated $P$. This can be partly attributed to the increased stochasticity of the switching current resulting from thermal fluctuations associated with higher microwave power. Indeed, it is more likely due to the suppression of $I_{sw}^{+}$ and $I_{sw}^{-}$ at elevated power $P$. When the switching current is smaller, identical fluctuation levels in $I_{sw}^{+}$ and $I_{sw}^{-}$ lead to more pronounced fluctuations in $\eta$, given that $\eta = \frac{\left|I_{sw}^{+}\right| - \left|I_{sw}^{-}\right|}{\left|I_{sw}^{+}\right| + \left|I_{sw}^{-}\right|}$. To further rule out the possibility of experimental artifacts, we directly sweep
$P$ while setting $B$ to $\Phi \textless$ 0.5$\Phi_{0}$ and $\Phi \textgreater$ 0.5$\Phi_{0}$ to measure the diode polarity switching, as illustrated in Fig.~\ref{fig4}(d). The diode polarity switching can be observed around -15dBm. Figure.~\ref{fig4}(c) presents a numerically simulated result using the RSJ model with previously obtained parameters. $i_{ac}$ are selected at values before reaching the first node of the zeroth Shapiro step lobe. The simulated result qualitatively agrees with the experimental observations. As $i_{ac}$ is tuned from 0.179$i_{c}$ to 1.786$i_{c}$, the sign of $\eta$ changes accordingly. The sign-reversal of the diode efficiency can be intuitively understood as follows:  In the fast-driving limit and considering the zero-voltage step, the amplitudes of the Josephson harmonics $K_{n}$ in the CPR are modulated by the zeroth Bessel function of the first kind~\cite{PhysRevResearch.6.L022002}. Specifically, $K_{1} \propto J_{0}(i_{ac}/\omega)$, $K_{2} \propto J_{0}(2i_{ac}/\omega)$. As $i_{ac}$ increases, both $K_{1}$ and $K_{2}$ can change signs, leading to a sign change of the diode efficiency. The microwave-controlled switching of diode polarity offers new possibilities for distinguishing 4$\pi$-periodic topological Josephson junctions in asymmetric SQUIDs\cite{PhysRevResearch.6.023011}.

\par In conclusion, we have developed a gate- and flux-tunable superconducting diode utilizing a SQUID composed of aluminum and InSb nanosheets.  Theoretical analyses indicate that higher harmonics dominate the SDE rather than $L$. Fractional Shapiro step experiments demonstrate the evolution of these higher harmonics with flux tuning, showing significant enhancements in the second-harmonic signatures of the SQUID near half-integer flux quanta, as confirmed by numerical simulations. Furthermore, by integrating numerical calculations with microwave experiments, we show that the diode polarity can be switched by tuning the microwave power in SQUIDs. These findings present an exciting perspective for exploring highly tunable superconducting diodes on nanosheet-based inteferometers and realizing a parity-protected qubit. They may also have potential applications in discerning 4$\pi$-periodic Josephson junctions.

\section{Author contributions}
H.Q.X. supervised the project. X.W., H.S., and S.Y. fabricated devices. X.W. and J.-Y.W. performed measurements. X.W. did the simulation. D.P. and J.Z developed the nanosheet material. X.W., P.Z., and H.Q.X. analyzed the data and wrote the manuscript with inputs from all the authors. 

\section{Acknowledgements}
This work is supported by the NSFC (Grant Nos. 92165208, 12004039, 11874071, 92365103, 12374480, 12374459, 61974138 and 92065106). D.P. acknowledges the support from Youth Innovation Promotion Association, Chinese Academy of Sciences (Nos. 2017156 and Y2021043). 

\section{Data analysis and data availability}
The raw data and the analysis files are available at \href{https://doi.org/10.5281/zenodo.13268561}{https://doi.org/10.5281/zenodo.13268561}

\bibliography{ref}

\end{document}


\title{Tunable superconducting diode effect in higher-harmonic InSb nanosheet interferometers}

\author{Xingjun Wu}
\email{wuxj@baqis.ac.cn}
\affiliation{Beijing Academy of Quantum Information Sciences, Beijing 100193, China}

\author{Ji-Yin Wang}
\affiliation{Beijing Academy of Quantum Information Sciences, Beijing 100193, China}
\author{Haitian Su}
\affiliation{Beijing Key Laboratory of Quantum Devices, Key Laboratory for the Physics and Chemistry of Nanodevices, and School of Electronics, Peking University, Beijing 100871, China}
\affiliation{Institute of Condensed Matter and Material Physics, School of Physics, Peking University, Beijing 100871, China}
\author{Shili Yan}
\affiliation{Beijing Academy of Quantum Information Sciences, Beijing 100193, China}
\author{Dong Pan}
\email{pandong@semi.ac.cn}
\affiliation{State Key Laboratory of Superlattices and Microstructures, Institute of Semiconductors,Chinese Academy of Sciences, P.O. Box 912, Beijing 100083, China}

\author{Jianhua Zhao}
\affiliation{State Key Laboratory of Superlattices and Microstructures, Institute of Semiconductors,Chinese Academy of Sciences, P.O. Box 912, Beijing 100083, China}
\author{Po Zhang}
\email{zhangpo@baqis.ac.cn}
\affiliation{Beijing Academy of Quantum Information Sciences, Beijing 100193, China}

\author{H. Q. Xu}
\email{hqxu@pku.edu.cn}
\affiliation{Beijing Academy of Quantum Information Sciences, Beijing 100193, China}
\affiliation{Beijing Key Laboratory of Quantum Devices, Key Laboratory for the Physics and Chemistry of Nanodevices, and School of Electronics, Peking University, Beijing 100871, China}

\maketitle

\newpage


\section{Loop inductance estimation}

\par The total loop inductance of the SQUID consists of the geometric inductance ($L_{geo}$) and the kinetic inductance ($L_{kin}$). $L_{geo}$ arises from the energy stored in the magnetic field due to the current caused by Faraday's laws, whereas $L_{kin}$ arises from the kinetic energy stored in the motion of the charge carriers. For a wire loop with area $S$ and circumference $l$, $L_{geo}$ can be approximated by the formula:
$$ L_{geo} = \frac{\mu_{0}l}{2\pi}\left(ln\frac{2S}{rl}-0.15+\frac{\zeta}{4}\right) $$
where $\mu_{0}$ is the magnetic vacuum permeability, $r$ is the wire radius, and $\zeta$ is a parameter related to the current distribution in the wire's cross-section. At low frequencies, $\zeta \approx$ 1. Our SQUID geometry gives $S \approx$ 70~$\mu$$m^{2}$, $r \approx$ 0.35~$\mu$m, and $l \approx$ 41~$\mu$m, thus we find $L_{geo} \approx$ 19.5~pH. 

The kinetic sheet inductance of an Al film, $L_{kin/\square}$, at temperatures $T \ll T_{c}$, can be evaluated using the zero-temperature superconducting gap $\Delta_{0}$ and the normal-state sheet resistance $R_{n/\square}$.
$$L_{kin/\square} = \frac{hR_{n/\square}}{2\pi^2\Delta_{0}}$$
Here, $\Delta_{0} = 1.76k_{B}T_{c}$, with $k_{B}$ being Boltzmann constant and $T_{c}$ the critical temperature of the Al film. Measurements of a 115~nm-thick Al film yield a normal-state sheet resistance $R_{n/\square} \sim$ 0.64 $\Omega$ and a critcial temperature $T_{c} \sim$ 1.2~K. Thus, $L_{kin/\square}$ = 0.74~pH. Considering approximately 68 squares in the SQUID, the estimated $L_{kin}$ is 50~pH. Consequently, the total inductance, $L_{tot}$, is estimated. To avoid underestimating the inductance, we assume $L_{tot}$ = 100~pH in the main text.  

\section{Device fabrication}

\par
\begin{figure}[H]\begin{center}
\includegraphics[width=1\linewidth]{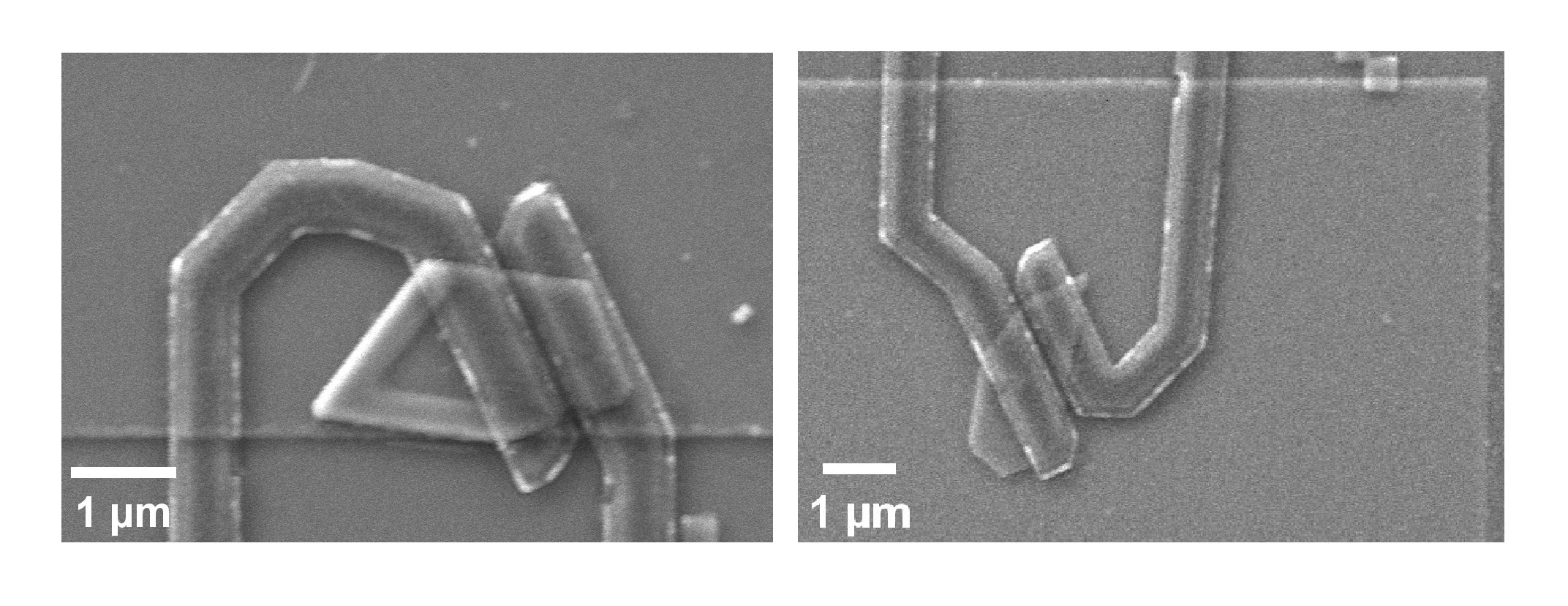}\end{center}
\caption{\label{fig:s1} 
The SEM images of JJ$_{1}$ (left) and JJ$_{2}$ (right) in our SQUID device.
}
\end{figure}

 First, we define two parallel strip patterns using electron beam lithography (EBL) on an $n$-doped Si/SiO$_{2}$ substrate, each strip being approximately 8~µm wide and spaced~12 µm apart. Subsequently, a metallic layer of 5~nm Ti/20 nm~Au is deposited to form two local backgates ($V_{bg1}$ and $V_{bg2}$), followed by a 15~nm HfO$_{2}$ dielectric layer applied via atomic layer deposition (ALD). After setting up the Ti/Au backgates, InSb nanosheets are carefully transferred onto each backgate using a tip under microscopic guidance. Suitable InSb nanosheets are selected using SEM for further device fabrication. Two types of PMMA (200K and 950K) are then spin-coated on the substrate and each baked at 170°C for 3 minutes. An SQUID structure, similar to the one shown in Fig. 1, is patterned using EBL and subsequently developed in a dedicated developer solution. To ensure a highly transmissive superconductor-semiconductor interface between aluminum and the InSb nanosheets, the exposed InSb surface is treated in a 40°C ammonium polysulfide solution (3 mol/L $(NH_{4})_{2}S_{x}$ : $H_{2}O$ = 1 : 9 by volume) for 2.5 minutes to remove any oxide layer. The sample is then quickly transferred to a thermal evaporation chamber within 2 minutes to prevent reoxidation, and a 5~nm Ti/115~nm Al layer is deposited. Following the aluminum deposition, the device undergoes a liftoff process in hot acetone. The resistance of the devices is measured at room temperature using a probe station, selecting those with resistance of 1 $\sim$ 10 k$\Omega$ for low-temperature measurements in a dilution refrigerator. Zoomed-in SEM images of JJ${1}$ and JJ$_{2}$ in the SQUID device described in the main text are shown in Fig.~\ref{fig:s1}.

\section{Superconducting diode effect at different gate voltages}
Figure S2 shows the diode effects at different $V_{bg2}$, when $V_{bg1}$ is fixed at 1.069~V. Each panel shows the four-probe measured voltage across the SQUID as a function of $I$ and normalized $\Phi/\Phi_{0}$. For each $\Phi/\Phi_{0}$, $I$ is swept from zero to positive values and then from zero to negative values. The red and blue branches correspond to the switching branches in opposite sweep directions. The switching current is asymmetric with respect to the current bias direction at $V_{bg2}$ = 1.819~V.  The switching valley in the positive bias (indicated by the red arrow) is located on the left shoulder of the negative bias (indicated by the blue arrow). As $V_{bg2}$ varies, the asymmetry of the switching current reverses, with the positive switching valley on the right shoulder of the negative one at $V_{bg2}$ = 0.9~V. This demonstrates an electrically tunable SDE in the SQUID.
\begin{figure}[H]
\includegraphics[width=1\linewidth]{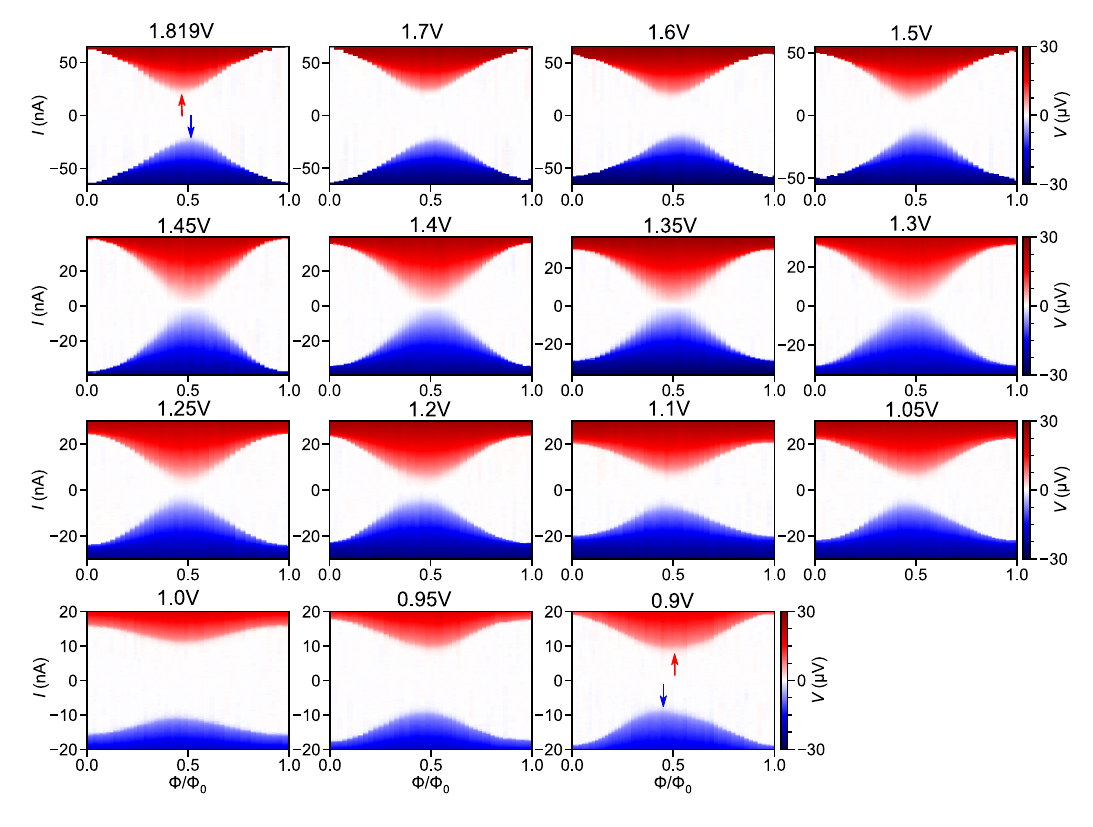}
\caption{ 
Electrically tunable diode effects in the SQUID. When tuning $V_{bg2}$ from 1.819~V to 0.9~V, the polarity of the diode reverses. Here,  $V_{bg1}$ is fixed at 1.069~V.
}
\label{fig:s2}
\end{figure}
~~~~~~~~~~~~~~~~~~~~~~~~~~~~~~~~~~~~~~~~~~~~~~~~~~~~~~~~~~~~~~~~~~~~~~~~~~~~~~~~~~~~~~~~~~~~~~~~~~~~~~~~~~~~~~~~~~~~~~~~~~~~~~~~~~~~~~~~~~~~~~~~~~~~~~~~~~~~~~~~~~~~~~~~~~~~~~~~~~~~~~~~~~~~~~~~~~~~~~~~~~~~~~~~~~~~~~~~~~~~~~~~~~~~~~~~~~~~~~~~~~~~~~~~~~~~~~~~~~~~~~~~~~~~~~~~~~~~~~~~~~~~~~~~~~~~~~~~~~~~~~~~~~~~~~~~~~~~~~~~~~~~~~~~~~~~~~~~~~~~~~~~~~~~~~~~~~~~~~~~~~~~~~~~~~~~~~~~~~~~~~~~~~~~~~~~~~~~~~~~~~~~~~~~~~~~~~~~~~~~~~~~~~~~~~~~~~~~~~~~~~~~~~~~~~~~~~~~~~~~~~~~~~~~~~~~~~~~~~~~~~~~~~~~~~~~~~~~~~~~~~~~~
\section{Numerical simulation: the effects of loop inductance and higher harmonic terms on the SDE of the SQUID}

\begin{figure}[H]\centering
\includegraphics[width=1\linewidth]{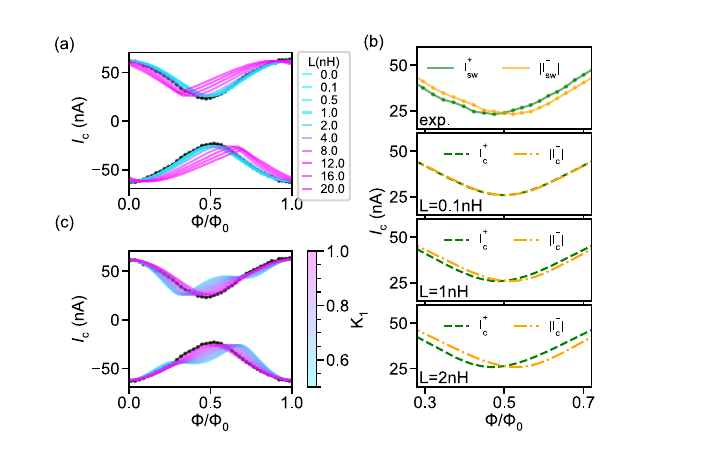}
\caption{\label{fig:s3} 
(a) Overall simulation of the effect of different inductances, $L$, on $I_{c}$, assuming the JJs composing the SQUID have a sinusoidal CPR. It is observed that as L increases, the asymmetry between $I_{c}^{+}$ and $I_{c}^{-}$ becomes more pronounced, and for larger $L$, the CPR exhibits a sawtooth-like shape.
(b) Detailed analysis of the simulated diode effect at $L$ = 0.1, 1 and 2~nH, assuming the JJ exhibits a CPR. The results indicate that for a realistic inductance of $L$ = 0.1~nH, the simulated diode effect is nearly imperceptible, which contradicts the experimental observations (see the top panel). To replicate the experimentally observed diode effect, the assumed inductance $L$ must be at least an order of magnitude larger than the realistic value, which is inconsistent with the physical reality. (c) Simulation of the impact of different higher-order harmonic components in the CPR on $I_{c}$. Here a realistic $L$ = 0.1 nH and the first two harmonics (the first harmonic:  $K_{1}$ and the second harmonic:  $K_{2}$ = 1-$K_{1}$) are considered. As the contribution from higher-order terms (1-$K_{1}$) increases, the asymmetry between $I_{c}^{+}$ and $I_{c}^{-}$ becomes more pronounced. The back dotted lines in (a) and (c) represent the experiment data at a gate configuration of ($V_{bg1}$, $V_{bg2}$) = (1.069 V, 1.819 V). Parameters ($I_{sw1}$, $I_{sw2}$) = (18 nA, 44 nA), obtained at this gate configuration, are used for all the simulation plots.}
\end{figure}
~~~~~~~~~~~~~~~~~~~~~~~~~~~~~~~~~~~~~~~~~~~~~~~~~~~~~~~~~~~~~~~~~~~~~~~~~~~~~~~~~~~~~~~~~~~~~~~~~~~~~~~~~~~~~~~~~~~~~~~~~~~~~~~~~~~~~~~~~~~~~~~~~~~~~~~~~~~~~~~~~~~~~~~~~~~~~~~~~~~~~~~~~~~~~~~~~~~~~~~~~~~~~~~~~~~~~~~~~~~~~~~~~~~~~~~~~~~~~~~~~~~~~~~~~~~~~~~~~~~~~~~~~~~~~~~~~~~~~~~~~~~~~~~~~~~~~~~~~~~~~~~~~~~~~~~~~~~~~~~~~~~~~~~~~~~~~~~~~~~~~~~~~~~~~~~~~~~~~~~~~~~~~~~~~~~~~~~~~~~~~~~~~~~~~~~~~~~~~~~~~~~~~~~~~~~~~~~~~~~~~~~~~~~~~~~~~~~~~~~~~~~~~~~~~~~~~~~~~~~~~~~~~~~~~~~~~~~~~~~~~~~~~~~~~~~~~~~~~~~~~~~~
\section{RSJ simulation: tuning superconducting diode response with AC currents at 8~GHz}

\begin{figure}[H]\centering
\includegraphics[width=1\linewidth]{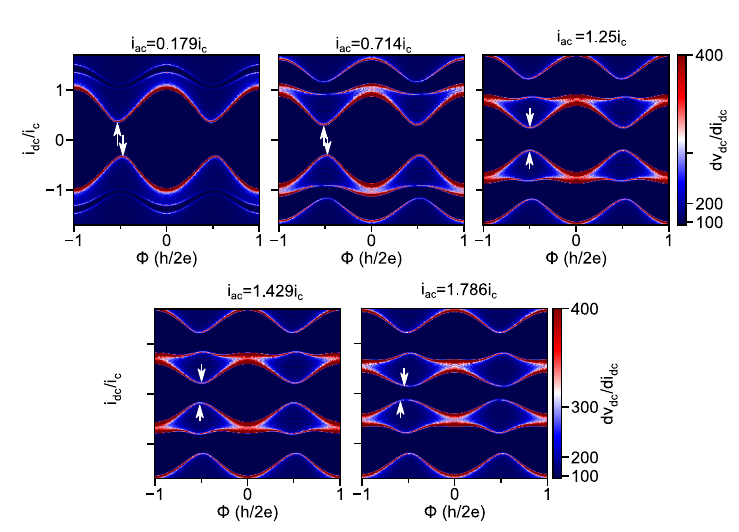}
\caption{\label{fig:s4} 
The calcultated differential resistance d$v_{dc}$/d$i_{dc}$ as a function of bias current $i_{dc}$ and magnetic flux $\Phi$ at different microwave amplitudes $i_{ac}$. The simulation parameters are R = 300~$\Omega$, microwave frequency $f$ = 8~GHz, $i_{c1}$ = 0.33$i_{c}$, $i_{c2}$ = 0.67$i_{c}$, and $K_{1}$ = 0.96. As $i_{ac}$ increases, the asymmetry of the critical current changes. The relative offset between the valleys in opposite directions shifts from positive to negative (see white arrows), indicating that the diode efficiency changes sign at larger $i_{ac}$.  
} 
\end{figure}

\section{Determination of switching currents at different gate voltages}
\par In the experiment, thermal and charge fluctuations inevitably introduce stochasticity in the switching current when varying $V_{bg}$. To simulate the data in Fig. 2(d), we use the Savitzky-Golay filter to extract the $V_{bg2}$-dependent switching current $I_{sw2, sing}$ in the single junction JJ$_{2}$, as shown by the solid line in Fig.~\ref{fig:s5}. Additionally, we note that the sum of $I_{sw1, sing}$ and $I_{sw2, sing}$ is consistently smaller than the switching current $I_{sq}$ through the SQUID at the same gate configuration. This occurs because the SQUID has a smaller resistance and lower dissipation, resulting in a larger $I_{sq}$. Therefore, we assume the current ratio between JJ$_{1}$ and JJ$_{2}$ is $\frac{I_{sw1, sing}}{I_{sw2, sing}}$ in our SQUID. Thus, $I_{sw1} =\frac{I_{sw1, sing}}{I_{sw1, sing}+I_{sw2, sing}}I_{sq}$ and $I_{sw2} =\frac{I_{sw2, sing}}{I_{sw1, sing}+I_{sw2, sing}}I_{sq}$. This method is used in the main text.

\begin{figure}[H]\centering
\includegraphics[width=0.7\linewidth]{figS5_backgate dependent ic.pdf}
\caption{\label{fig:s5} 
Switching current in in JJ$_{2}$ versus backgate voltage $V_{bg2}$ at $V_{bg1}$ = 0~V.  
} 
\end{figure}

\section{Evolution of Shapiro island under different gate configurations}
In the presence of microwave power with $P$ = -12.5~dBm, $f$ = 6~GHz, we measured the evolution of the Shapiro island as $V_{bg2}$ is tuned from 1.7~V to 1.0~V, while keeping $V_{bg1}$ = 1.069~V fixed. As $V_{bg2}$ decreases, the zero voltage gap narrows first and then widens. When tuning $V_{bg2}$ close to 1.3~V, the critical currents satisfy $I_{sw2} \approx I_{sw1}$. Under this condition, the zero voltage gaps at half-integer flux quanta become difficult to clearly resolve due to the small higher harmonic content in the CPR. Additionally, it is noteworthy that under the condition of $I_{sw2} \approx I_{sw1}$, we did not observe the integer Shapiro steps with the same step height as the fractional Shapiro steps at $\Phi$ = $\Phi_{0}$/2, as theoretically expected in the symmetric case. Instead, the fractional Shapiro steps dominated, while the integer steps became significantly suppressed. Furthermore, at $\Phi$ = $\Phi_{0}$/4, the expected disappearance of the half-integer steps was not observed.

We believe the deviation from the expected results is primarily due to the fact that the CPRs of the two junctions are not exactly the same, even though they have the same critical currents. This arises because the SQUID consists of two independent InSb nanosheets, each fabricated into a Josephson junction through ex-situ fabrication techniques. The inherent differences between the two InSb nanosheets, along with the instability in the ex-situ fabrication process during the preparation of the superconducting-semiconductor interface, lead to varying transparency between the two junctions. As a result, the CPRs of the junctions differ.

In the context of studying symmetric SQUIDs, the simple assumption of identical CPRs for each junction in the form $I_{i}=\alpha I_{ci} (K_{1} sin\varphi+K_{2} sin2\varphi)$, where $K_{1}$ and $K_{2}$ are constants, is no longer appropriate. The difference in the values of ($K_{1}$, $K_{2}$) between the junctions leads to the $sin\varphi$ term in the SQUID CPR no longer disappearing when $\Phi$ = $\Phi_{0}$/2. Similarly, at $\Phi$ = $\Phi_{0}$/4, the $sin2\varphi$ term still remains. However, this simplified assumption remains sufficiently valid for the qualitative analysis of asymmetric SQUIDs.

\begin{figure}[H]\centering
\includegraphics[width=1\linewidth]{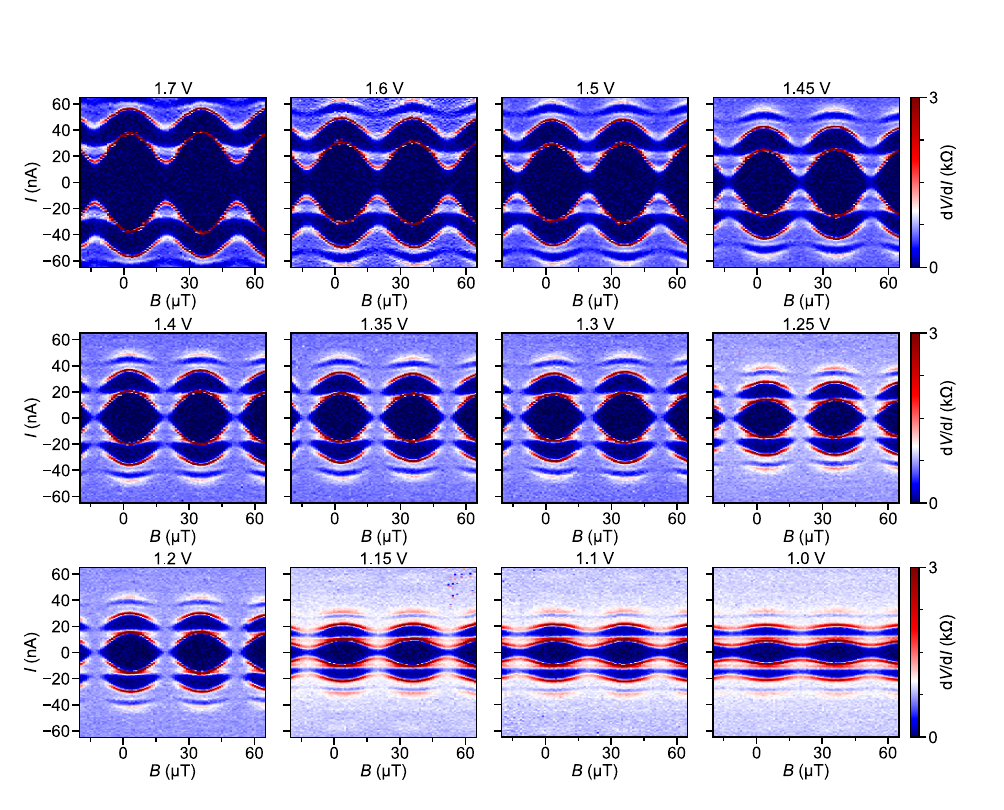}
\caption{\label{fig:s6} 
The evolution of the Shapiro island when $V_{bg2}$ is tuned from 1.7~V to 1.0~V. Here, $P$ = -12.5~dBm, $f$ = 6~GHz, and $V_{bg1}$ = 1.069~V.} 
\end{figure}

\section{Enhanced asymmetry by increasing $\left|\frac{K_{2}}{K_{1}}\right|$}

\begin{figure}[H]\centering
\includegraphics[width=0.9\linewidth]{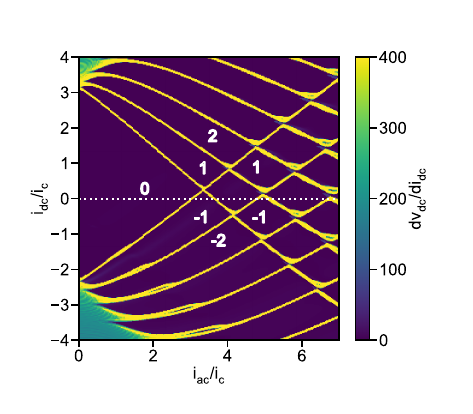}
\caption{\label{fig:s7} 
Differential resistance, d$v_{dc}$/d$i_{dc}$, versus $i_{dc}$ and $i_{ac}$ at $\Phi/\Phi_{0}$ = 0.25, showing a strong asymmetry around $i_{dc}$ = 0. Here ($K_{1}$, $K_{2}$) = (0.5, 0.5). Other parameters simulated for the plot are the same as in Fig.3 of the main text. It can be observed that the increase in higher-order harmonic components is a crucial factor in generating significant asymmetry.
}
\end{figure}

\bibliography{}